\documentclass[11pt]{article}
\usepackage{latexsym,amssymb}

\begin{document}

\def\bea{\begin{eqnarray}}
\def\eea{\end{eqnarray}}
\def\be{\begin{equation}}
\def\ee{\end{equation}}
\def\nn{\nonumber\\}

\title{Quantised bulk fields in the Randall-Sundrum compactification model\footnote{To appear in Physics Letters B.}}

\author{David J.~Toms\thanks{E-mail address:
{\tt d.j.toms@newcastle.ac.uk}}\\
Department of Physics, University of Newcastle upon Tyne,\\
Newcastle Upon Tyne, United Kingdom NE1 7RU}

\date{May 1, 2000}
\maketitle
\begin{abstract}The quantisation of a scalar field in the five-dimensional model suggested by Randall and Sundrum is considered. Using the Kaluza-Klein reduction of the scalar field, discussed by Goldberger and Wise, we sum the infinite tower of modes to find the vacuum energy density. Dimensional regularisation is used and we compute the pole term needed for renormalisation, as well as the finite part of the energy density. Some comments are made concerning the possible self-consistent determination of the radius.
\end{abstract}
\eject

For about the past year there has been a resurgence of interest in the old Kaluza-Klein idea that spacetime may have some extra dimensions.  One reason for this renewed interest lies in the solution it provides to the hierarchy between the Planck scale and the electroweak scale \cite{first}.  In place of the compact extra dimensions used in \cite{first}, Randall and Sundrum \cite{RS} have proposed a 5-dimensional model in which the extra dimension has an orbifold compactification with two 3-branes with opposite tensions sitting at two orbifold fixed points.  An essential feature of this model is that the 5-dimensional geometry is not a direct product of the 4-dimensional spacetime and the extra dimension.  The line element, which is a solution to the 5-dimensional Einstein equations for a particular 3-brane source, is
\be
ds^2 = e^{-2kr|\phi|}\eta_{\mu \nu} dx^\mu dx^\nu -r^2d\phi^2\;.\label{1}
\ee
Here $x^\mu $ are the usual coordinates on 4-dimensional spacetime, and $ |\phi |\le\pi$ with the points $(x^\mu,\phi) $ and $(x^\mu, -\phi) $ identified.  The 3-branes sit at $\phi = 0,\pi$.  $k$ is a constant and $r$ is also constant and associated with the size of the extra dimensions.  The constant $r$ is an arbitrary constant of integration found when solving the Einstein equations. Randall and Sundrum showed \cite{RS} that for $kr\sim10$, essentially because of the exponential appearing in (\ref{1}), that TeV mass scales could be produced in the 4-dimensional theory from Planck scale quantities in the higher-dimensional theory.  Furthermore the Randall-Sundrum model has possible  phenomenological consequences that put higher-dimensional models within the experimental range.  (See also \cite{third}.)

 In the Randall-Sundrum model, the radius $r$ of the extra dimension is not fixed.  This was also the case for the older Kaluza-Klein theories which were based on the simpler geometry of a direct product of 4-dimensional Minkowski spacetime and a homogeneous space, often chosen to be a sphere.  It was realised by Candelas and Weinberg \cite{CW} that quantum effects from matter fields, or gravity, could be used to fix the size of the extra dimensions and obtain a self-consistent solution.  In view of the interest generated by the Randall-Sundrum scenario, it is reasonable to ask if the radius of the extra dimension in their model can be determined in a similar way.  This is the prime motivation for the present paper in which we calculate the vacuum (or Casimir) energy of a quantised scalar field on the Randall-Sundrum background, since this is a first step in all self-consistency calculations.  As we will see, the orbifold nature of the Randall-Sundrum spacetime makes this calculation considerably more difficult than was the case in the original Kaluza-Klein models based on direct product spacetimes.

It was shown in \cite{GW1} that a scalar field on the higher dimensional spacetime could be considered, from the 4-dimensional point of view, in the usual Kaluza-Klein way as an infinite tower of scalar fields whose masses are quantised. The action can be written as
\be
S=\frac{1}{2}\sum_{n}\int d^4x\left(\partial^\mu\varphi_n\partial_\mu\varphi_n-m_n^2\varphi_n^2\right)\;,\label{2}
\ee
where $m_n^2$ gives the mass for the $n^{th}$ mode. All dependence on the extra fifth dimension has been integrated out here to leave a 4-dimensional theory. For the Randall-Sundrum model (\ref{1}) Goldberger and Wise \cite{GW1} showed that the masses $m_n$ are given by solutions to the transcendental equation
\be
0=y_\nu(ax_n)j_\nu(x_n)-j_\nu(ax_n)y_\nu(x_n)\label{3}
\ee
where we have defined $a=e^{-\pi kr}$ and $m_n=kax_n$, with $x_n$ the $n^{th}$ positive solution to (\ref{3}). The functions $j_\nu$ and $y_\nu$ are shorthand for the following combinations of Bessel functions~:
\bea
j_\nu(z)&=&2J_\nu(z)+zJ_\nu'(z)\label{4a}\\
y_\nu(z)&=&2Y_\nu(z)+zY_\nu'(z)\label{4b}\;.
\eea
The order $\nu$ of the Bessel functions is given by
\be
\nu=\sqrt{4+\frac{m^2}{k^2}}\;,\label{5}
\ee
where $m$ is the mass of the 5-dimensional scalar field. The derivation of all of the results (\ref{2}--\ref{5}) is described very clearly in \cite{GW1}.  The fact that $a = e^{-\pi kr}\sim10^{-17} $, which is necessary for a solution to the hierarchy problem \cite{RS}, and that as a consequence $x_1\sim1$ allows the conclusion \cite{GW1} that the lightest modes have masses of the order of a few TeV.  This is in marked contrast to the older Kaluza-Klein picture in which a whole spectrum of light and unobserved particles could appear.

What we wish to do in this paper is calculate the quantum vacuum polarisation effects from the infinite tower of massive particles in (\ref{2}).  As mentioned earlier, similar calculations were performed in the older Kaluza-Klein models based on direct products of Minkowski spacetime with homogeneous spaces, such as spheres (see for example \cite{CW,oldKK} and references therein) with the aim of obtaining self-consistent solutions for the size of the extra dimensions.  It seems a natural progression of this earlier work to ask if similar calculations can be done in the new Randall-Sundrum model.  A complication which is present in this model is that it is not possible to obtain a closed form expression for the mass modes $m_n$, unlike the case of direct products of flat spacetime with homogeneous spaces.\footnote{Of course it is trivial to obtain approximate solutions to (\ref{3}) using the asymptotic form of the Bessel functions of large argument, but these are useless for our purposes.}  Nevertheless we will describe a procedure for evaluating the vacuum energy using only the basic properties of (\ref{3}) irrespective of the lack of an explicit form for $m_n$.

In this paper we will concentrate on the one-loop vacuum energy density for the theory whose Lagrangian is given by (\ref{1}) which is
\bea
V^{(1)} & = & \frac{\ell^\epsilon} {2}\sum_n\int \frac{ d^{3 +\epsilon} }{(2\pi)^{3 +\epsilon}} (p^2 + m_n^2)^{1/2}\nn
& = & \frac{ 1} {16\pi^2}\left (\frac{\ell^2} {4\pi}\right)^{\epsilon/2}\Gamma (-2-\epsilon/2)\sum_{n = 1}^ {\infty} (m_n^2)^{2 +\epsilon/2}\;.\label{6}
\eea
Here we are using dimensional regularisation with $\ell$ the renormalisation length.  ($\zeta $-function regularisation can be used to the same end.)  The $\epsilon\rightarrow 0$ limit is understood here, but before this limit is taken it is necessary to evaluate the sum over modes in (\ref{6}).  Since $m_n^2 = (ka)^2x_n^2$, we need to evaluate
\be
v (\epsilon) =\Gamma (-2-\epsilon/2)\sum_{n = 1}^{\infty} x_n^{4 +\epsilon}\;,\label{7}
\ee
with $x_n$ the $n ^{th} $ positive solution to (\ref{3}).  Then
\be
V^{(1)} = \frac{ (ka)^4} {16\pi^2}\left (\frac{ k^2a^2\ell^2} {4\pi}\right)^{\epsilon/2} v (\epsilon)\;.\label{8}
\ee
will give the quantum vacuum energy (in the limit $\epsilon\rightarrow0$).

 The sum defining $v (\epsilon) $ can be evaluated in the usual way by first converting it into a contour integral in the region of the complex $\epsilon $-plane where the sum converges (which is easily seen to be $\Re\epsilon<-5$ in this case) and then deforming the contour so that we can perform an analytic continuation back to a neighbourhood of $\epsilon = 0$.  Similar calculations have been done for the Casimir effect in a spherical shell \cite{shell}.  It is possible to show that
\be
v (\epsilon) = \frac{ 1} {\Gamma (3 +\epsilon/2)}\int_{0}^{\infty} dz\, z^{4 +\epsilon} \frac{ d} {dz}\ln P_\nu (z)\;,\label{9}
\ee
where
\be
P_\nu (z) = \frac{ 2} {\pi}\left\lbrack k_\nu (z) i_\nu (az) -k_\nu (az) i_\nu (z)\right\rbrack\;.\label{10}
\ee
$k_\nu (z) $ and $i_\nu (z) $ are given by expressions like (\ref{4a}) and (\ref{4b}) but involve the modified Bessel functions $K_\nu (z) $ and $I_\nu (z) $ in place of $J_\nu (z) $ and $Y_\nu (z) $.

The remaining task now is to perform the analytic continuation of $v (\epsilon) $ to a region about $\epsilon = 0$.  This can be done by studying the behaviour of the integrand at both small and large $z$.  The analytic continuation is most easily performed by splitting up the integration range in (\ref{9}) and defining
\be
v (\epsilon) = v_1 (\epsilon) + v_2 (\epsilon)\;,\label{11}
\ee
where
\be
v_1 (\epsilon) =\frac{ 1} {\Gamma (3 +\epsilon/2)}\int_{0}^{1} dz\, z^{4 +\epsilon} \frac{ d} {dz}\ln P_\nu (z)\;,\label{12}
\ee
 with $v_2 (\epsilon) $ given by a similar expression but with the integration going from 1 to $\infty$.  Use of the small $z$ behaviour of the Bessel functions  shows that $v_1 (\epsilon) $ is analytic at $\epsilon = 0$.  This means that we may concentrate on $v_2 (\epsilon) $ whose analytic continuation is less trivial.  The basic idea is to study how the integral defining $v_2 (\epsilon) $ diverges as $\epsilon\rightarrow 0$, and by adding and subtracting appropriate terms enable the $\epsilon\rightarrow 0$ limit to be taken everywhere apart from a possible pole term.

The impediment to the convergence of $v_2 (\epsilon) $ at $\epsilon = 0$ is the behaviour of the integrand for large $z$.  We can use the known asymptotic behaviour of the Bessel functions to calculate the terms we need to add and subtract to let $\epsilon= 0$.   We will define
\bea
i_\nu (z) & = &\sqrt{\frac{ z} {2\pi}} e^z\Sigma_\nu (z)\;,\label{13a}\\
k_\nu (z) & = &-\sqrt{\frac{ \pi z} {2}} e^{-z}\Sigma_\nu (-z)\;,\label{13b}
\eea
with $\Sigma_\nu (z)\sim1$ as $z\rightarrow\infty$.  This means that the second term in $P_\nu (z) $ is the most divergent (see (\ref{10})) as $z\rightarrow\infty$.  We find
\bea
v_2 (\epsilon) & = & -\frac{ 1} {8} -\frac{ 1} {10} (1-a) + \frac{ 1} {2}\int_{1}^{\infty} dz\, z^4 \frac{ d} {dz}\ln\left\lbrack 1-\frac{ k_\nu (z) i_\nu (az)} {i_\nu (z) k_\nu (az)}\right\rbrack\nn
& & + \frac{ 1} {\Gamma (3 +\epsilon/2)}\int_{1}^{\infty} dz\, z^{4 +\epsilon} \frac{ d} {dz}\ln\left\lbrack\Sigma_\nu (z)\Sigma_\nu (-az)\right\rbrack + {\mathcal O} (\epsilon)\;.\label{14}
\eea
To obtain this result it is necessary to assume $\Re\epsilon < -5$, and perform an analytic continuation to $\epsilon = 0$.  The first integral in (\ref{14}) is easily seen to be finite, and we only need to take care with the second integral.  Define, for large $z$,
\be
\ln\Sigma_\nu (z)\sim\sum_{k = 1}^{\infty} \frac{ d_k} {z^k}\label{15}
\ee
for some coefficients $d_k$.  By adding and subtracting the first five terms in the asymptotic expansion we can end up with a result which is finite at $\epsilon = 0$.  It is easy to see that
\bea
v_2 (\epsilon) & = &  -\frac{ 1} {8} -\frac{ 1} {10} (1-a) + \frac{ 1} {2}\int_{1}^{\infty} dz\, z^4 \frac{ d} {dz}\ln\left\lbrack 1-\frac{ k_\nu (z) i_\nu (az)} {i_\nu (z) k_\nu (az)}\right\rbrack\nn
&&+ \frac{ 1} {2}\int_{1}^{\infty} dz\, z^{4 } \frac{ d} {dz}\left\lbrace\ln\left\lbrack\Sigma_\nu (z)\Sigma_\nu (-az)\right\rbrack-\sum_{k=1}^{5}\left(1+\frac{(-1)^k}{a^k}\right)\frac{d_k}{z^k}\right\rbrace\nn
&&+\frac{1}{6}\left(1-\frac{1}{a}\right)d_1+\frac{1}{2}\left(1+\frac{1}{a^2}\right)d_2+\frac{3}{2}\left(1-\frac{1}{a^3}\right)d_3\nn
&&+\frac{4}{\epsilon\Gamma(3+\epsilon/2)}\left(1+\frac{1}{a^4}\right)d_4-\frac{5}{2}\left(1-\frac{1}{a^5}\right)d_5\;.\label{16}
\eea
The coefficients $d_k$ defined in (\ref{15}) are given in Table~1. The only task remaining is the evaluation of the two integrals, which must be done numerically.

\begin{table}[t]
\begin{center}
\renewcommand{\arraystretch}{1.5}
\begin{tabular}{|c|c|}\hline
$d_1$&$\frac{1}{8}(13-4\nu^2)$\\ \hline
$d_2$&$\frac{1}{16}(4\nu^2-19)$\\ \hline
$d_3$&$\frac{1}{384}(16\nu^4-200\nu^2+481)$\\ \hline
$d_4$&$-\frac{1}{128}(16\nu^4-104\nu^2+187)$\\ \hline
$d_5$&$-\frac{1}{5120}(64\nu^6-880\nu^4+5228\nu^2-9029)$\\ \hline
\end{tabular}\end{center}
\caption{Coefficients $d_k$ defined in (\ref{15}) for $k=1,\ldots,5$.}
\end{table}

Because $d_4\ne0$ the vacuum energy density contains a pole term as $\epsilon\rightarrow 0$.  The origin of this divergence lies in the orbifold nature of the Randall-Sundrum spacetime.  Without the orbifold fixed points the result for the vacuum energy density would have been finite (after regularisation).  Similar divergences have been discussed on manifolds with conical singularities \cite{cones} and arise in calculations involving black hole entropy.  In our case the pole part of $V^{(1)} $ is given by
\be
{\rm P.P.}\lbrace V^{(1)}\rbrace=-\frac{(1+a^4)}{1024\pi^2\epsilon}(16m^4+24m^2k^2+25k^4)\;.\label{17}
\ee
This result has used the expression for $d_4$ in Table~1, and the definition of $\nu$ in (\ref{5}). If we had used $\zeta$-function regularisation, in place of the above pole term we would have found a non-zero result for $\zeta(0)$ given by an expression like (\ref{17}). Even if we use the fact that $a<<1$, there is still a pole term present in the vacuum energy density with a non-zero coefficient. It is also noteworthy that the pole term is present even if $m=0$ corresponding to an initially massless scalar field. 

 The evaluation of the remaining parts of the vacuum energy involves numerical calculations.  There is no impediment to doing these calculations for general values of $m, k, a$; however, because the main purpose of the present paper is just to illustrate the general method, we will make life simpler for ourselves.  If, following \cite{GW1}, we are interested in $m\sim k\sim1$ (in Planck units) we can take $m/k = 3/2$ leading to (from (\ref{6})) $\nu = 5/2$.  This simplifies all of the Bessel functions to powers and exponentials:
\bea
k_{5/2}(z) & = & -\frac{ 1} {4}\sqrt {2\pi}\; z^{-5/2} (2z^3 + 3z^2 + 3z + 3)e^{-z}\;,\label{18a}\\
i_{5/2} (z) & = & (2\pi)^{-1/2} z^{-5/2}\left\lbrack (2z^3 + 3z)\cosh z-3 (1 + z^2)\sinh z\right\rbrack\;.\label{18b}
\eea
The other complication present is the extreme smallness of the parameter $a$; but because all integrals are now finite it is possible to expand in powers of $a$ in a straightforward way and obtain the leading order behaviour (in $a$) of the finite part of $V^{(1)}$, which we call ${\rm F.P.}\lbrace V^{(1)}\rbrace$. After some calculation we find
\bea
{\rm F.P.}\lbrace V^{(1)}\rbrace&=&\frac{k^4}{16\pi^2}\Big\lbrace-\frac{81}{64}\ln(k^2\ell^2)+5.84858058+\frac{15}{16}a-\frac{3}{8}a^2\nn
&&-a^4\Big\lbrack\frac{81}{64}\ln(k^2a^2\ell^2)-2.293923\Big\rbrack+{\mathcal O}(a^5)\Big\rbrace\;.\label{19}
\eea

This completes our evaluation of the vacuum energy density. Before it can be used to examine the problem of self-consistency for the Randall-Sundrum spacetime there are several important things that remain to be done. The first is to perform a careful renormalization analysis. It should be possible to do this following analogous calculations in the case of conical singularities \cite{cones}. A second thing that is essential to consider is the effect of the induced gravity term which will arise, as first discussed in \cite{TomsPLB}, and whose importance in self-consistency was noted by Candelas and Weinberg \cite{CW}. Ideally it will be possible to perform the analogous calculations for quantum gravity and study the role of self-consistency in the Randall-Sundrum model. (Some work on the Kaluza-Klein reduction of fields of spin other than zero has been done in \cite{GP}.) Whether or not it is possible to fix the value of the arbitrary compactification radius $r$ by such a calculation, and provide an alternative to the calculation in \cite{GW2}, remains to be seen. Further aspects of the calculations will be given elsewhere \cite{FlachiToms}.

\vspace{1cm}\noindent{\bf Acknowledgements}:~I would like to thank A. Flachi for helpful discussions, and J.~S.~Dowker for supplying some references for heat kernels on cones.

\end{document}